\documentclass[sort&compress
    ,final            
  ]
  {aipproc}

\layoutstyle{8x11single}

\usepackage{amssymb}

\newcommand{\unit}[1]{\mathrm{#1}}
\newcommand{\keV}{\unit{keV}}
\newcommand{\GeV}{\unit{GeV}}
\newcommand{\TeV}{\unit{TeV}}
\newcommand{\km}{\unit{km}}
\newcommand{\sqcm}{\unit{cm}^{2}}
\newcommand{\percubcm}{\unit{cm}^{-3}}
\newcommand{\persqkm}{\unit{km}^{-2}}
\newcommand{\kmpers}{\unit{km/s}}
\newcommand{\pers}{\unit{s}^{-1}}
\newcommand{\peryr}{\unit{yr}^{-1}}

\begin{document}
\rightline{\normalfont{\small ULB-TH/09-27; DESY 09-193 }}
\title{Neutrino signature of Inert Doublet Dark Matter}

\classification{95.35.+d, 12.60.Fr, 95.85.Ry} %
\keywords      {dark matter, astronomical observations of theories
beyond the SM, neutrino experiments}

\author{Sarah Andreas}{
    address={Institut f\"ur Theoretische Physik E,
    RWTH Aachen University, D-52056 Aachen, Germany}
    ,altaddress={Service de Physique Th\'eorique,
    Universit\'e Libre de Bruxelles, B-1050 Brussels, Belgium \\
    Deutsches Elektronen-Synchrotron DESY, Notkestra{\ss}e 85, D-22607
    Hamburg, Germany\footnote{Currently at DESY, email: Sarah.Andreas@desy.de}}}

\begin{abstract}
In the framework of the Inert Doublet Model and extensions, the
signature of neutrinos from dark matter annihilation in the Earth,
the Sun and at the Galactic centre is presented. The model contains
an extra Higgs doublet, a neutral component of which is chosen as
dark matter candidate. There are three distinct mass ranges for
which consistency both with WMAP abundance and direct searches can
be obtained: a low (4 - 8 GeV), a middle (60 - 70 GeV) and a high
\mbox{(500 - 1500 GeV)} WIMP mass range. The first case is of
interest as we showed that the model can at the same time give the
correct WMAP abundance and account for the positive DAMA results
without contradicting other direct searches. We present how capture
in the Sun can further constrain this scenario using
Super-Kamiokande data. Indirect detection through neutrinos is
challenging for the middle and high mass ranges. For the former, the
presence of the so-called `iron resonance' gives rise to larger
neutrino fluxes for WIMP masses around 60 - 70 GeV since capture by
the Earth is enhanced. The addition of light right-handed Majorana
neutrinos to the particle content of the model further increases the
signal since it opens a direct annihilation channel into
mono-energetic neutrinos. Neutrinos from the Galactic centre might
be detected for heavy WIMPs if the dark matter density at the
Galactic centre is substantially boosted.
\end{abstract}

\maketitle


\section{Introduction}

Driven by observations indicating that most matter in the universe
is non-luminous, various dark matter candidates have been proposed.
The candidate of the Inert Doublet Model (IDM) studied in this work
belongs to the often studied category of weakly interacting massive
particles (WIMPs)~\cite{Jungman:1995df,Bertone:2004pz}. Introduced
in~\cite{Deshpande:1977rw} as a simple standard model extension with
an additional Higgs doublet and a scalar dark matter candidate, the
IDM provides despite its simplicity a quite rich phenomenology. This
has already been emphasized in several dedicated
articles~\cite{Ma:2006km,Barbieri:2006dq,Hambye:2007vf},
investigations regarding LHC~\cite{Dolle:2009ft} and studies of the
specific dark matter signatures which have been carried out for
direct detection~\cite{Andreas:2008xy,Arina:2009um} as well as
indirect detection in gamma
rays~\cite{LopezHonorez:2006gr,Gustafsson:2007pc},
neutrinos~\cite{Andreas:2009hj,Agrawal:2008xz} and
antimatter~\cite{Nezri:2009jd}. An interpretation of the IDM in a
grand unified theory framework is given in~\cite{Kadastik:2009cu}.

\begin{figure}[hbt!]
\includegraphics[clip = true, viewport = 5.1cm 17.1cm 15.7cm 25.4cm, width=7.5cm]{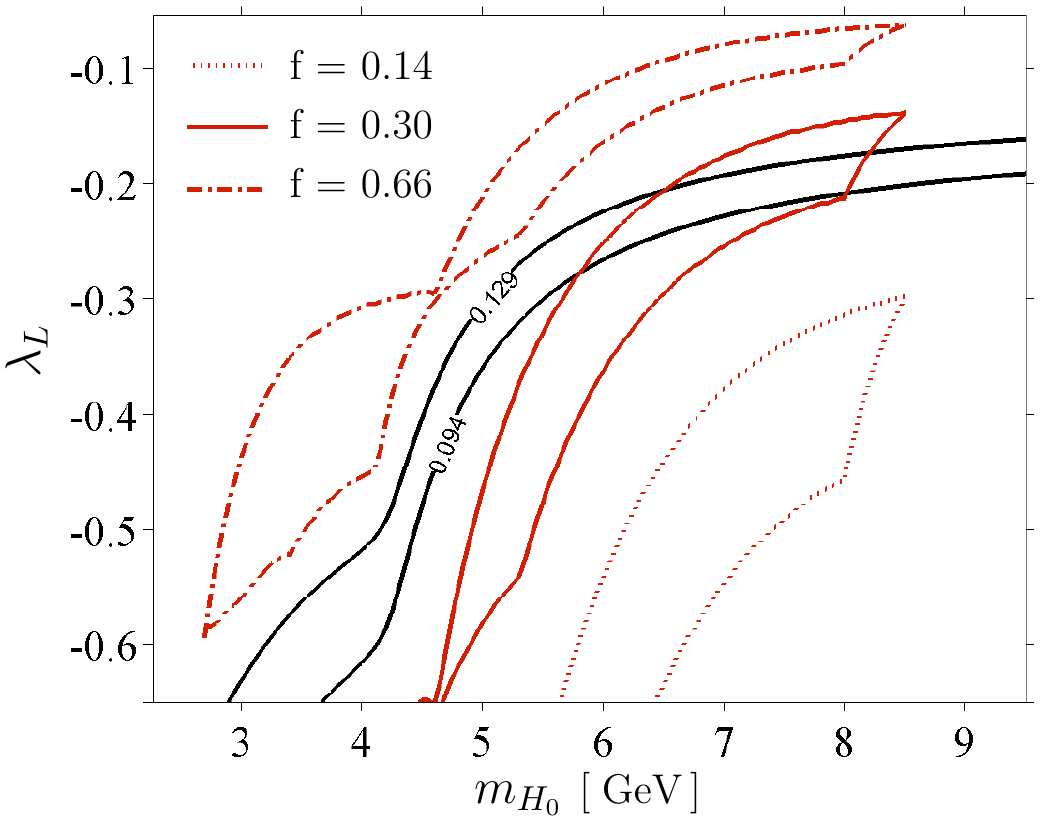}
\hspace{0.2cm}
\includegraphics[clip = true, viewport = 2.1cm 16.5cm 19cm 27.2cm, width=8.5cm]{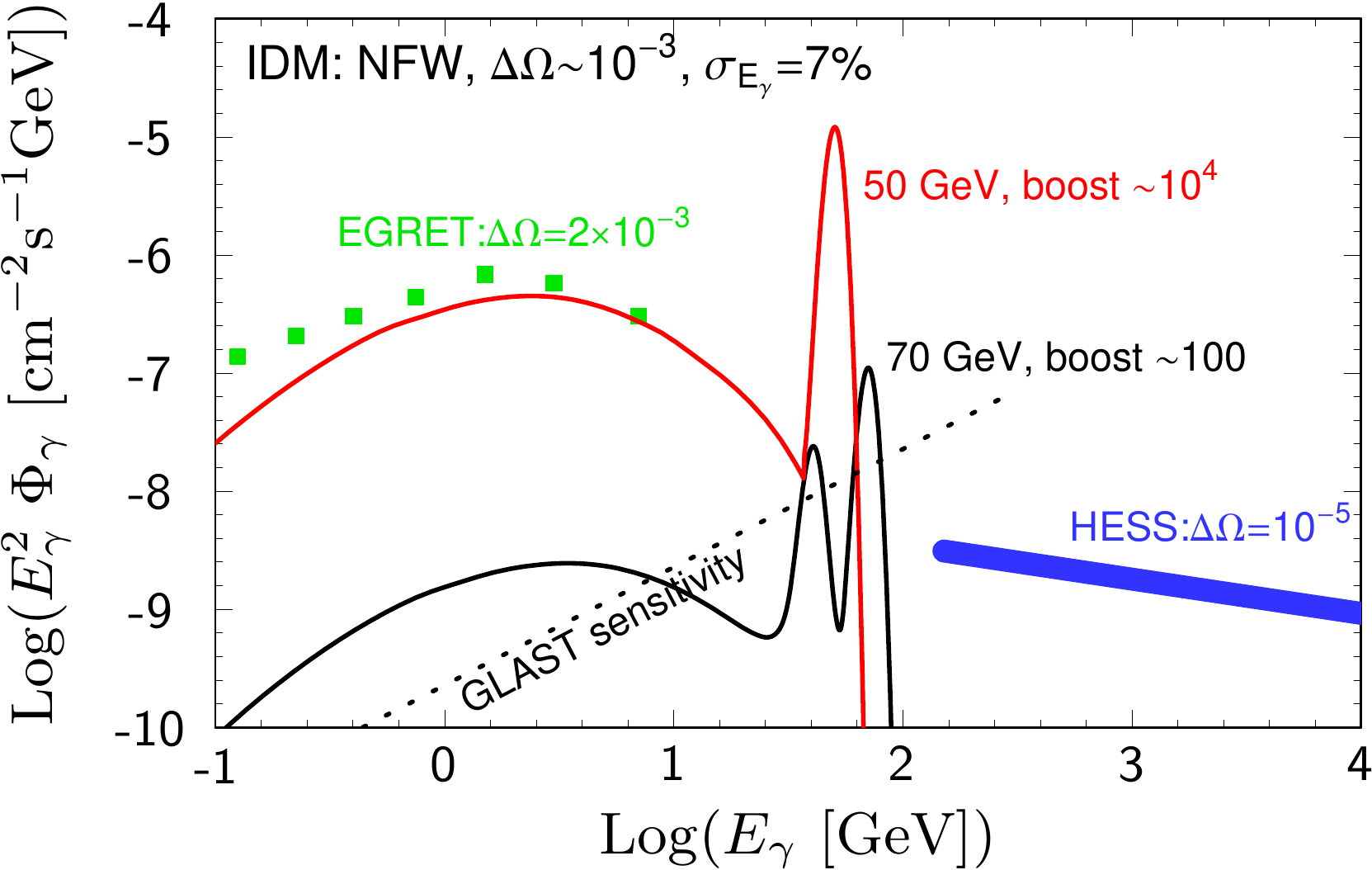}
\caption{Features of the Inert Doublet model in view of recent or
upcoming experiments: \newline \textit{left:} Regions between the
red lines can account for the DAMA annual modulation signal and
nicely overlap with the one in which the correct WMAP abundance is
obtained (between the black lines)~\cite{Andreas:2008xy}. \newline
\textit{right:} Gamma lines from the annihilation of inert doublet
dark matter into $\gamma\gamma$ and $\gamma Z$ (at loop-level) at
the Galactic centre might be observed with the Fermi satellite whose
sensitivity is shown as dotted line~\cite{Gustafsson:2007pc}.
\label{fig-DAMA-Gamma}}
\end{figure}

In figure~\ref{fig-DAMA-Gamma}, we emphasize two examples of
interesting features connected with recent or upcoming observations.
It has been shown in~\cite{Petriello:2008jj,Savage:2008er} that
light WIMPs with spin-independent scattering on nuclei can account
for the annual modulation signal reported by the DAMA
collaboration~\cite{Bernabei:2008yi} without contradicting the null
results of the other direct detection experiments. In the framework
of the IDM, it was found in~\cite{Andreas:2008xy} that agreement
with DAMA can be obtained within the red regions of
figure~\ref{fig-DAMA-Gamma} (left plot) drawn in function of the
dark matter mass $m_{H_0}$ and the only other relevant parameter of
the model $\mu_2$ (the three regions result from the uncertainty on
the Higgs to nucleus coupling $f$). Additionally, the abundance lies
within the WMAP range between the two black lines, resulting in an
overlap where both DAMA and WMAP can be fulfilled at the same time.
Another test of this interesting scenario is possible with neutrinos
from the Sun as will be discuss in this work. The right plot in
figure~\ref{fig-DAMA-Gamma} shows a ``smoking gun'' signal of
monochromatic gamma-rays from the Galactic centre which has been
presented in~\cite{Gustafsson:2007pc}. Those gamma lines result from
a direct annihilation of the IDM dark matter candidate into
$\gamma\gamma$ and $\gamma Z$, possible at loop-level and give a
clear signal that might be observed with the upcoming Fermi
satellite. The right relic density is obtained for those models due
to a significant amount of coannihilations with the other inert
particle $A_0$ of the model in the early Universe.

\section{The Inert Doublet Model (IDM)}

The IDM is a standard model extension with two Higgs doublets
($H_1$, $H_2$) and an unbroken $Z_2$ symmetry under which
\begin{equation}
H_1 \rightarrow H_1 \;\; \mathrm{and} \;\; H_2\rightarrow - H_2
\end{equation}
while all the other Standard Model particles are even. Its potential
can be written as
\begin{equation}
\label{potential} V ~=~ \mu_1^2 \vert H_1\vert^2 + \mu_2^2 \vert
H_2\vert^2  + \lambda_1 \vert H_1\vert^4 + \lambda_2 \vert
H_2\vert^4 + \lambda_3 \vert H_1\vert^2 \vert H_2 \vert^2 +
\lambda_4 \vert H_1^\dagger H_2\vert^2 + \frac{\lambda_5}{2}
\left[(H_1^\dagger H_2)^2 + h.c.\right],
\end{equation}
where $H_1$ contains the Standard Model Higgs particle $h$. The
discrete symmetry prevents flavour changing neutral currents and
guarantees the stability of the lightest odd state. The lighter of
the neutral components of the extra doublet $H_2 = (H^+, 1/\sqrt{2}
(H_0 +i A_0))^T$ is therefore a candidate for dark matter and was in
this work chosen to be $H_0$. We define $\lambda_L=(\lambda_3 +
\lambda_4 + \lambda_5)/2$ as the coupling between $h$ and a pair of
$H_0$ and use as input parameters $\mu_2$, $\lambda_2$ together with
the masses of the scalar particles, including the Higgs. We chose
the parameters to be in agreement with the constraints discussed
in~\cite{LopezHonorez:2006gr} and experimental bounds
from~\cite{Barbieri:2006dq,Lundstrom:2008ai,Cao:2007rm} and shade in
the following figures those regions which are excluded by either of
the two.

Under the assumption that $H_0$ was in thermal equilibrium in the
early universe agreement of the abundance of $H_0$ (which we compute
using micrOMEGAs~\cite{Belanger:2006is}) with WMAP is found for a
low ($3 \ \GeV \lesssim m_{H_0} \lesssim 8 \ \GeV$), middle ($40 \
\GeV \lesssim m_{H_0} \lesssim 80 \ \GeV$) and high mass ($500 \
\GeV \lesssim m_{H_0} \lesssim 1500 \ \GeV$) range in agreement
with~\cite{LopezHonorez:2006gr,Dolle:2009fn}. For each of those mass
ranges a different annihilation channel is dominant. A light $H_0$
annihilates through the Higgs channel into fermions. In addition to
this, if kinematically allowed, coannihilation with $A_0$ through a
$Z$ boson takes place for the middle mass range. Annihilation into
$W^\pm$ pairs opens above 80 GeV and brings the abundance below WMAP
because of its large cross section. Even though the annihilation
cross section decreases and the relic abundance increases for higher
masses, there is a parameter region consistent with WMAP around 1
TeV~\cite{LopezHonorez:2006gr}.

A process relevant for the later discussion is the elastic
scattering of $H_0$ on a nucleus $\mathcal{N}$ which is
spin-independent in the framework of the IDM and occurs through
Higgs exchange. Its low energy cross section is given by
\begin{equation}
\sigma_{H_0\mathcal{N}_i\rightarrow H_0\mathcal{N}_i}^{SI} ~ = ~
\frac{1}{\pi} ~ \frac{\lambda_L^2}{m_h^4} ~
\frac{m_{\mathcal{N}_i}^4}{(m_{H_0} +  m_{\mathcal{N}_i})^2} ~ f^2
\, , \label{eq-sigmaH0NSI}
\end{equation}
where $m_{\mathcal{N}_i}$ is the mass of the nucleus and $f$ is the
nuclear form factor parameterizing the Higgs to nucleon coupling.
This factor is related to the trace anomaly $f m_{\mathcal{N}}\equiv
\langle \mathcal{N}| \sum_q m_q \bar{q}q|\mathcal{N}\rangle =
g_{h\mathcal{NN}} v$ but is unfortunately not well known. From the
results quoted in~\cite{Andreas:2008xy}, we take for a general
nucleon $f=0.30$ as central value from the rather wide range $0.14 <
f < 0.66$.

\section{Indirect Detection}

Dark matter can be detected indirectly through the observation of
its annihilation products among which the focus of this article lies
on neutrinos. In general, the highest fluxes of annihilation
products are expected to come from places with high dark matter
density and accordingly high annihilation rate. Dark matter can
accumulate over time in gravitational wells like the Galactic centre
or get captured in astronomical bodies like the Sun or the Earth.
Experiments like IceCube, Antares and Super-Kamiokande then search
for secondary neutrinos from dark matter annihilation. Depending on
the neutrino energy spectrum one distinguishes two extreme cases
referred to as soft ($\nu$ from $b\overline{b}$) and hard ($\nu$
from $W$ and $Z$) spectra (the spectra of the different annihilation
channels are computed with WimpSim~\cite{Wimpsim,Blennow:2007tw}).
Detectors are more sensitive to the latter since a hard spectrum is
flatter and has significant contributions at high energies. In the
following, we present an overview of the relevant processes and
formulas we used to obtain the results that will be presented in the
next section.

\subsection{Dark matter annihilation at the Galactic centre}

A large amount of dark matter is expected to be concentrated around
the Galactic centre thus opening the possibility for indirect
detection through its annihilation products. Those fluxes of
secondary particles strongly depend on the assumed dark matter
density profile. Even though indications for a high dark matter
density are also found in numerical simulations, the exact shape of
this density profile especially close to the Galactic centre is
still poorly known. At the Galactic centre, a flat behaviour is
expected for a cored profile from observations of rotation
curves~\cite{Flores:1994gz,Kravtsov:1997dp}, while a more cuspy
shape seems to be preferred by numerical simulations (see Kravtov et
al.~\cite{Kravtsov:1997dp}, Navarro-Frenk-White
(NFW)~\cite{Navarro:1995iw} and Moore et al.~\cite{Moore:1999nt} as
well as the recent Via Lactea and AQUARIUS
simulations~\cite{Diemand:2008in,Springel:2008cc}). Due to this
uncertainty boost factors are often considered in the analysis,
expressing the fact that for higher dark matter densities larger
fluxes can be expected. For instance, in~\cite{Gustafsson:2007pc},
the predicted gamma flux is boosted by a factor $\sim 10^4$.

In addition to this astrophysical boost, the annihilation cross
section for heavy dark matter candidates that interact through
relatively lighter particles (for instance, the $W$, the $Z$ and the
Higgs in the present case) might for low relative velocities be
larger than the one at tree-level. This effect, called ``Sommerfeld
enhancement''~\cite{Hisano:2004ds,Cirelli:2007xd}, might be
non-negligible and was found in models similar to the IDM like
Minimal Dark Matter to be of ${\cal O}(10^2)$ at
most~\cite{Cirelli:2007xd,cirelli-2008}.

Despite those possibilities of boosting the flux, agreement with
existing observational constraints for example in gamma rays must be
maintained. Therefore, we follow a procedure proposed
in~\cite{bertone-2004-70} to circumvent the astrophysical
uncertainties and to obtain an upper limit on the neutrino flux that
might be reached either by a steeper dark matter profile and/or a
Sommerfeld enhancement. The gamma ray spectrum from the Galactic
centre is constrained by the observations of the EGRET
satellite~\cite{Hunter1997}. Requiring the predicted gamma flux to
stay below the observed one at all energies (an energy of $E_\gamma
= 10 \ \GeV$ is found to be the most constraining), a maximal boost
factor for a NFW profile is obtained for each point of the parameter
space. The resulting boosts of $\sim 10^2 - 10^3$ are quite moderate
considering that Sommerfeld enhancement might apply in addition to
the astrophysical uncertainties. The maximum allowed neutrino flux
is then computed according to
\begin{equation}
\phi_\nu^{max} ~ = ~ \phi_\nu^{NFW} \ \frac{\phi_\gamma^{EGRET}(E =
10\ \GeV)}{\phi_\gamma^{NFW}(E = 10 \ \GeV)} \label{eq-GCNumax}.
\end{equation}

\subsection{Capture and annihilation in the Sun/Earth}

Over time, astronomical bodies like the Sun or the Earth might also
obtain a large dark matter density at their centre by accumulating
WIMPs that are captured after having lost a part of their energy in
elastic scattering on a nucleus. As discussed in the previous
section, the coupling to nuclei is spin-independent in the IDM and
the cross section depends on the mass of the nuclei, cf.
equation~(\ref{eq-sigmaH0NSI}). Since composition and size of Sun
and Earth are different so is their ability to capture dark matter.
The Sun is big and mainly composed out of Hydrogen which due to its
small mass makes spin-dependent interactions with $\sigma \propto
J(J+1)$ preferable. However, in the case of spin-independent
interactions light particles are also well captured in the Sun. The
much smaller Earth consists dominantly of heavy nuclei and therefore
capture is strong for spin-independent interactions.

Our calculation of the capture rate is based on an elementary
treatment and follows the simple approximation
of~\cite{Jungman:1995df,Press:1985ug} which is expected to work for
the Sun but not necessarily for the Earth since the escape velocity
of the Earth is small relative to the mean velocity in the halo
making capture in the Earth sensitive to the (not well understood)
small velocity part of the distribution. In~\cite{Jungman:1995df},
the Earth is treated as if it was in free space, while in general
the gravitational field of the Sun would have to be taken into
account. However, it has been found that the approximations
of~\cite{Jungman:1995df} work reasonably well for dark matter masses
close to the iron resonance\footnote{Detailed investigations of this
subject can be found
in~\cite{Gould:1987ww,Gould:1999je,Damour:1998rh,Lundberg:2004dn,Peter:2009mi,Peter:2009mm}.},
which is just the interest of this work.

The capture rate in the Sun/Earth for spin-independent scattering on
nuclei is then given by
\begin{equation} \label{eq-CaptureRate}
C^{\odot/\oplus} \; = ~ c_{\odot/\oplus} ~ \frac{\rho_{local}}{0.3 \
\GeV \ \percubcm} ~ \frac{270 \ \kmpers}{\overline{v}_{local}} ~
\sum_i \, F_i \, f_i \, \phi_i \, S_i \,
\frac{\GeV^2}{m_{\mathcal{N}_i} \, m_{H_0}} \,
\frac{\sigma_i}{10^{-40} \ \sqcm},
\end{equation}
where $c_\odot ~ = ~ 4.8 \, * \, 10^{24} \ \pers$ ($c_\oplus ~ = ~
4.8 \, * \, 10^{15} \ \pers $) sets the scale for WIMP capture by
the Sun (Earth), cf. ref.~\cite{Jungman:1995df} and $\sigma_i$ has
to be taken from equation~(\ref{eq-sigmaH0NSI}). The local dark
matter density and mean velocity are set to the typical values
$\rho_{local} = 0.3 \ \GeV$ and $\overline{v}_{local}=270 \
\kmpers$. In the summation over all elements in the Sun/Earth, we
use the values given in~\cite{Jungman:1995df} for the mass fraction
$f_i$ and the distribution $\phi_i$ of element $i$. The kinematics
of the collision are encoded in the nuclear form factors $F_i =
F_i(m_{H_0})$ and the suppression factor $S_i =
S_i\big(\frac{m_{H_0}}{m_{N_i}}\big)$, both defined
in~\cite{Jungman:1995df}, where also a detailed discussion of the
behaviour of the capture rate as a function of the dark matter mass
can be found.

In the following, we will consider dark matter with a mass close to
the one of iron thereby taking advantage of the so called iron
resonance which enhances the capture rate in the Earth. For details
about the origin of this and other similar resonances please refer
to~\cite{Andreas:2009hj,Jungman:1995df}. From simple kinematics, it
is known that the energy loss of a WIMP is largest and therefore
capture most efficient if it scatters on a nucleus with similar
mass. Since iron is abundant in the Earth, the capture rate is large
for WIMPs with masses around 60 GeV.

Once captured, dark matter annihilates at the centre of the
Sun/Earth producing a flux of secondary particles $i$ given by
\begin{equation}
\Big(\, \frac{d\phi}{dE} \, \Big)_i ~ = ~ \frac{1}{2} ~ \frac{C \,
F_{\mathrm{EQ}}}{4 \pi R^2} ~ \sum_F ~ BR_F ~ \Big(\, \frac{dN}{dE}
\, \Big)_{F,i} \ , \label{eq-NuFluxSE-CFEQ}
\end{equation}
where $C$ is the capture rate, cf. equation~(\ref{eq-CaptureRate}),
$R$ is either the distance from the Sun to the Earth ($R \, = \,
d_{\odot-\oplus} \, = \, 1.5 * 10^8 \ \km$) or the radius of the
Earth ($R \, = \, r_{\oplus} \, = \, 6300 \ \km$) and $F_{EQ}$ is a
factor defined in~\cite{Jungman:1995df} describing whether or not
equilibrium between capture and annihilation has been established.
The sum is carried out over all kinematically allowed final states
$F$ with branching ratio $BR_F$ and differential spectrum
$(dN/dE)_{F,i}$ of annihilation products $i$.

Neutrinos from dark matter annihilation at the Sun or the Earth can
be observed with neutrino telescopes such as IceCube or
Super-Kamiokande. In general the best directional information is
obtained for muonic neutrinos by detection of the muon which is
produced in a charged-current interactions in the rock below the
detector. Therefore, in the following, we focus on muon-neutrinos
and present the detection rate of neutrino-induced, through-going
muon events from the Sun or the Earth given by
\begin{equation} \label{eq-MuFluxSE}
\phi_\mu^{\odot/\oplus} ~ = ~ \, \varphi_{\odot/\oplus} ~~
\frac{C^{\odot/\oplus} \, F_{\mathrm{EQ}}}{2 \ \pers} ~ \Big( \,
\frac{m_{H_0}}{\ \GeV} \, \Big)^2 ~ \sum_i \, a_i \, b_i ~ \sum_F \,
BR_F \, \langle \, N z^2 \rangle_{F,i} \, (m_{H_0}),
\end{equation}
where $\varphi_\odot ~ = ~ 2.54 \, * \, 10^{-23} \ \persqkm \
\peryr$ and $\varphi_\oplus ~ = ~ \varphi_\odot ~ ( d_{\odot-\oplus}
\, / r_{\oplus} )^2  ~ = ~ \varphi_\odot ~ * ~ 5.6 \, * \, 10^{6}$,
cf.~\cite{Jungman:1995df}. The neutrino-scattering coefficients
$a_i$ and muon-range coefficients $b_i$ are given
in~\cite{Jungman:1995df} as $a_\nu \, = \, 6.8$, $a_{\overline{\nu}}
\, = \, 3.1$, $b_\nu \, = \, 0.51$ and $b_{\overline{\nu}} \, = \,
0.67$. The second moment of the muon-neutrino spectrum $\langle \, N
z^2 \rangle_{F,i}(m_{H_0})$ scaled by the square of the injection
energy $E_{in}$ is
\begin{equation}
\langle \, N \, z^2 \, \rangle_{F,i} \,(E_{in}) ~ = ~
\frac{1}{E_{in}^2} \, \int \, \Big( \, \frac{dN}{dE} \, \Big)_{F,i}
\, (E_{\nu}, E_{in}) \, E_{\nu}^2 \, dE_{\nu}, \label{eq-IntNuSpec}
\end{equation}
where the integration contains as lower bound the threshold energy
of the detector.

\section{IDM neutrino signatures}

In this section, our results on the neutrino fluxes from the
annihilation of inert doublet dark matter are presented. From the
discussion of the previous sections follows that in each of the
three mass ranges where the abundance is in agreement with WMAP it
is most interesting to study neutrinos from one specific origin. For
heavy dark matter candidates, capture in the Sun or the Earth is
suppressed because of the $1/m_{DM}^2$ dependence of the capture
cross section~(\ref{eq-sigmaH0NSI}) designating the Galactic centre
as the most promising place to observe. Dark matter candidates of
the middle range have a mass close to the iron resonance and
therefore experience a strong enhancement of their capture rate in
the Earth. The lightest IDM candidates do not benefit from
resonances but are best captured in the Sun due to the $1/m_{DM}^2$
dependence favouring light particles.

\subsection{Galactic centre and heavy dark matter}
The most promising signal from the annihilation of dark matter in
the mass range $500 \ \GeV \lesssim m_{H_0} \lesssim 1500 \ \GeV$ is
expected to come from the Galactic centre. For such heavy $H_0$
candidates, it has been argued in ref.~\cite{LopezHonorez:2006gr}
that only $m_{H_0} \sim \mu_2$ gives the correct WMAP abundance and
is in agreement with model constraints (vacuum stability and
perturbativity). In this region of the parameter space, annihilation
occurs dominantly into $W$ and $Z$ bosons, both producing hard
neutrinos.

\begin{figure}[hbt!]
\includegraphics[clip = true, viewport = 1.1cm 11.3cm 20.2cm 26.4cm, width=8.5cm]{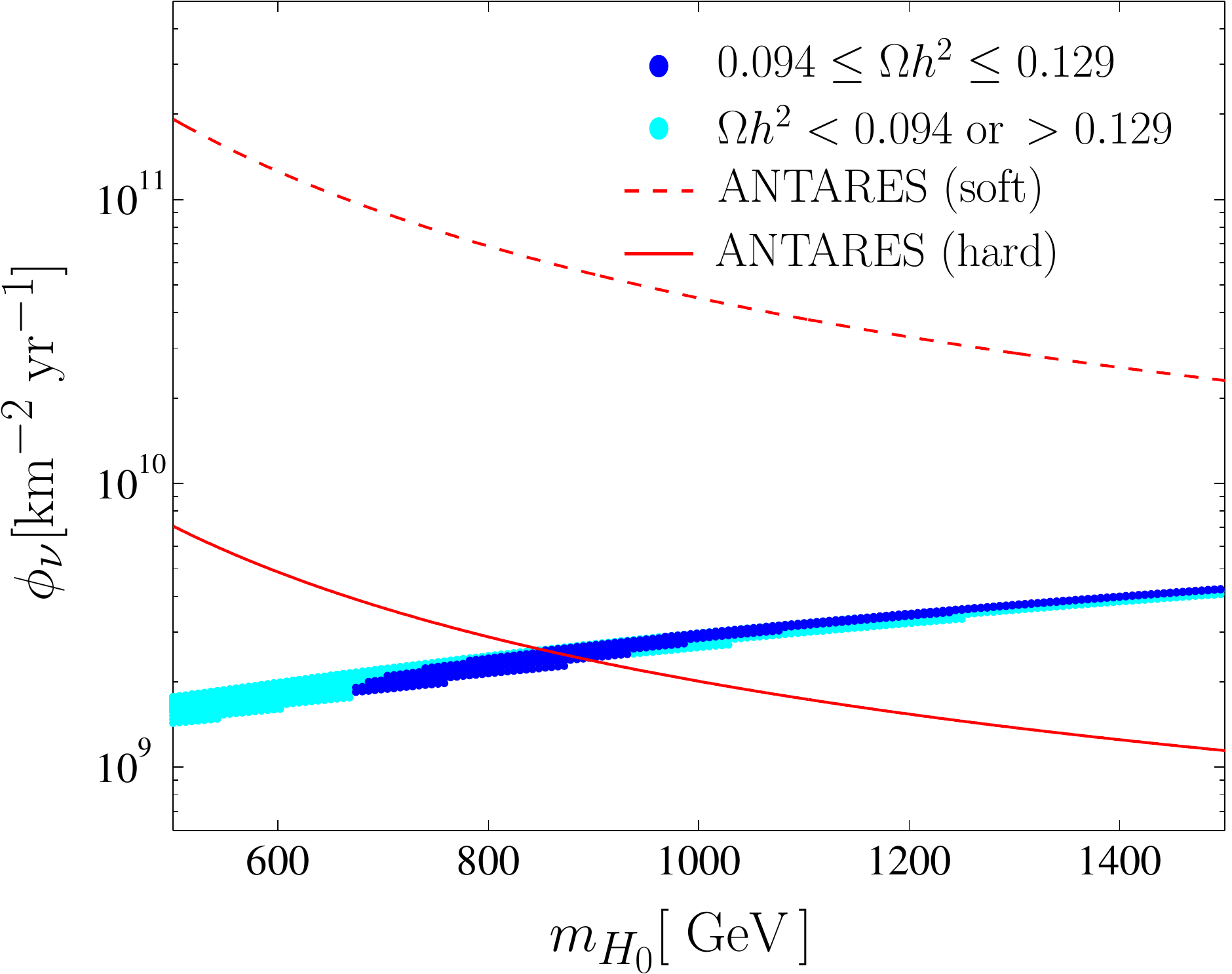}
\caption{Scatter plot of the maximum expectable neutrino flux in the
Antares detector for the IDM for which the corresponding
$\gamma$-flux does not exceed the one observed by EGRET. The
sensitivities of Antares for a soft (dashed) and a hard (solid)
spectrum are shown in red. The dark blue points lie within the WMAP
area while the light blue ones are outside. (Parameters: $m_h = 120
\ \GeV$,  $\lambda_2 = 0.2$, $\Delta m_{A_0 H_0} = 5 \ \GeV$,
$\Delta m_{H^+ H_0} = 10 \ \GeV$). \label{fig-GCnumaxflux}}
\end{figure}

In figure~\ref{fig-GCnumaxflux}, we show the upper limit on the
expected neutrino flux in Antares from the annihilation of $H_0$ at
the Galactic centre computed according to
equation~(\ref{eq-GCNumax}). All points in the scatter plot are
consistent with model constraints. The dark blue ones agree within
$2\sigma$ with WMAP while the light blue ones give too high or too
low abundance. The sensitivity of Antares taken
from~\cite{Bailey2002} is given both for a hard (solid red line) and
a soft (dashed red line) neutrino spectrum. Since the neutrinos
result mainly from $W$ and $Z$ decay their spectrum can be
considered hard and should be compared to the solid sensitivity
line. The maximum flux then lies above the Antares sensitivity for
$H_0$ masses above 900 GeV and might potentially lead to a signal.
For several models in this range, the dark matter abundance is in
agreement with WMAP.

\subsection{Earth and iron mass range}

\begin{figure}[hbt!]
\includegraphics[clip = true, viewport = 2.1cm 21.1cm 16.9cm 27.6cm, width=16.4cm]{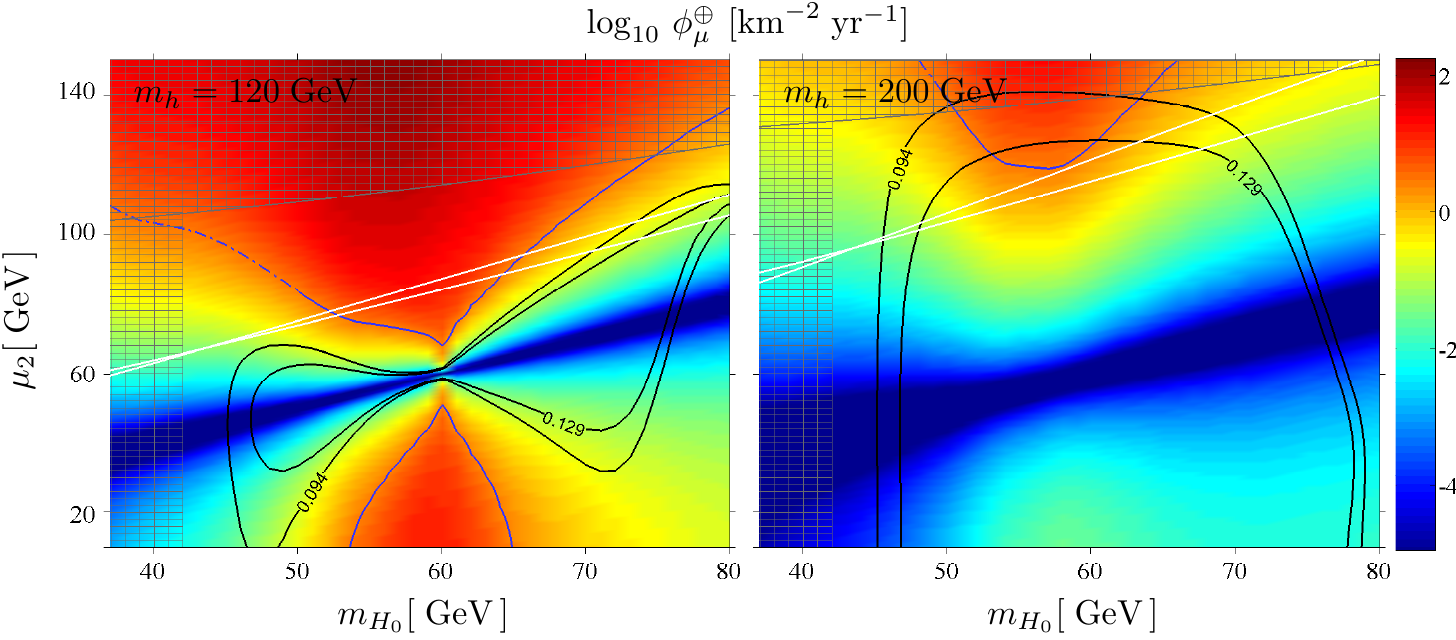}
\caption{Upper estimate of the muon flux in IceCube (without
detector threshold) from annihilation at the Earth. Colour gradient
- $\, \log_{10} \phi_\mu ~ [\persqkm \ \peryr]$; WMAP area - black
lines; XENON, CDMS limits - white lines; IceCube sensitivity soft -
blue lines; excluded regions - shaded. (Parameters: $\lambda_2 =
0.2$, $\Delta m_{A_0 H_0} = 8 \ \GeV$, $\Delta m_{H^+ H_0} = 50 \
\GeV$, $f = 0.3$) \label{fig-Earthbm}}
\end{figure}

As discussed earlier, the Earth captures efficiently dark matter
with a mass close to the one of iron thus providing an interesting
place to search for a signal. An upper estimate on the expected muon
events in IceCube obtained from equations~(\ref{eq-MuFluxSE})
and~(\ref{eq-IntNuSpec}) without applying detector thresholds is
plotted in figure~\ref{fig-Earthbm} in function of the dark matter
mass $m_{H_0}$ and the bare mass scale $\mu_2$ for two different
Higgs masses. The shaded parts are excluded, the black lines give
the $2\sigma$ WMAP region and the blue ones the sensitivity of
IceCube. In the considered dark matter mass range, neutrinos come
mainly from $b\overline{b}$ and have to be compared to the
sensitivity for a soft neutrino spectrum. Direct detection (white
lines, XENON and CDMS~\cite{collaboration-2008}) strongly constrains
the parameter space and excludes all regions for which the correct
abundance and a flux above the sensitivity is obtained. Our results
are in agreement with those of~\cite{Agrawal:2008xz}.

In the following, we discuss two possibilities that might improve
the prospects for IceCube. Recent numerical simulations indicated
the presence of a disc consisting of dark matter in our
Galaxy~\cite{Read:2008fh,Bruch:2008rx}. Since the relative velocity
of WIMPs in the dark disc compared to the solar system is smaller
then for usual WIMPs from the halo, the capture rate in the Earth
would be enhanced. The authors find in their study which is
performed in the framework of the MSSM that the muon fluxes from the
Earth might then be boosted by a factor 10-100~\cite{Bruch:2009rp}.
A similar enhancement applied to the IDM gives a more promising
prediction for the muon fluxes in IceCube.

An extension of the model with three right-handed Majorana neutrinos
that has been proposed in~\cite{Ma:2006km} in order to generate
neutrino masses through radiative corrections is also interesting in
terms of neutrino fluxes. It opens a new annihilation channel for
$H_0$ directly into neutrinos as shown in figure~\ref{fig-EarthMaj2}
thereby producing a monochromatic flux. The IDM Lagrangian must be
enlarged by the new terms
\begin{equation}
\mathcal{L} \supset  \, h_{ij} \, ( \nu_i \, H_0 - l_i \, H^+ )N_j +
{1\over 2} M_j  N_j N_j + h.c.\, \label{eq-MajYukawa}
\end{equation}
where the index $i$ stands for the SM neutrino generation ($i = e,
\mu, \tau$) and $j$ for the three extra Majorana neutrinos ($j =
1,2,3$). Even though the Majorana neutrinos are also odd under $Z_2$
we keep $H_0$ as dark matter candidate and find
\begin{equation}
\sigma |\vec{v}| ~ = ~ \frac{h^4}{4 \pi} ~ \frac{m_N^2}{(m_{H_0}^2 +
m_N^2)^2} \label{eq-sigmaH0Majnu}
\end{equation}
for its annihilation cross section into $\nu\nu$ or the conjugate
channel $\overline{\nu}\overline{\nu}$. For simplicity, we focus on
muon-neutrinos and suppose that only one Majorana neutrino, with
mass $m_N$, enters the cross section. Also we do not take into
account neutrino oscillations, which is a good approximation for the
Earth. We take the Majorana neutrino to be rather light ($m_N = 100
\ \GeV$) and $h = 0.1$ in order to get a large neutrino signal.

\begin{figure}[hb!]
\includegraphics[clip = true, viewport = 2.1cm 19.1cm 16.9cm 25.6cm, width=16.4cm]{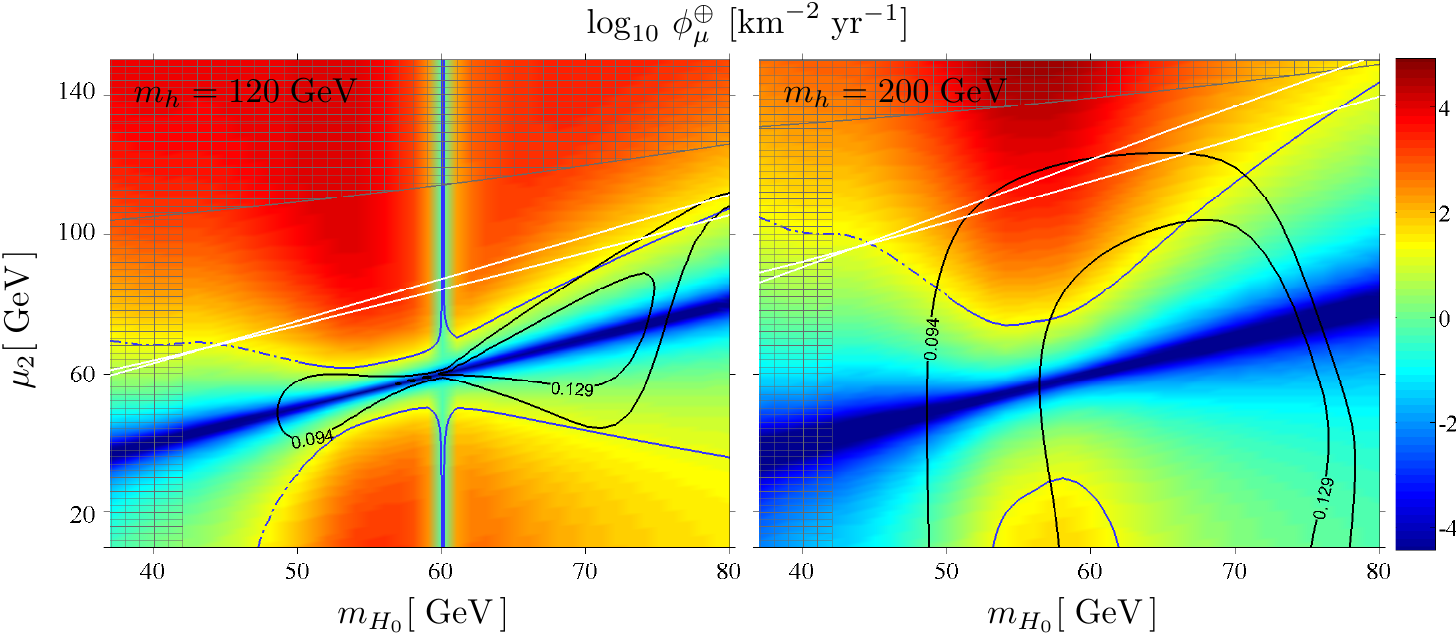}
\caption{Muon flux from the Earth for the IDM extended with light
Majorana neutrinos opening annihilation into mono-energetic
neutrinos. Colour gradient - $\, \log_{10} \phi_\mu ~ [\persqkm \
\peryr]$; WMAP area - black lines; XENON, CDMS limits - white lines;
IceCube sensitivity hard - blue lines; excluded regions - shaded.
(Parameters: $\lambda_2 = 0.2$, $\Delta m_{A_0 H_0} = 8 \ \GeV$,
$\Delta m_{H^+ H_0} = 50 \ \GeV$, $m_N = 100 \ \GeV$, $h = 0.1$, $f
= 0.3$).\label{fig-EarthMaj1}}
\end{figure}

\begin{figure}[hbt!]
\includegraphics[clip = true, viewport = 2.1cm 11.0cm 15.5cm 18.4cm, width=12.5cm]{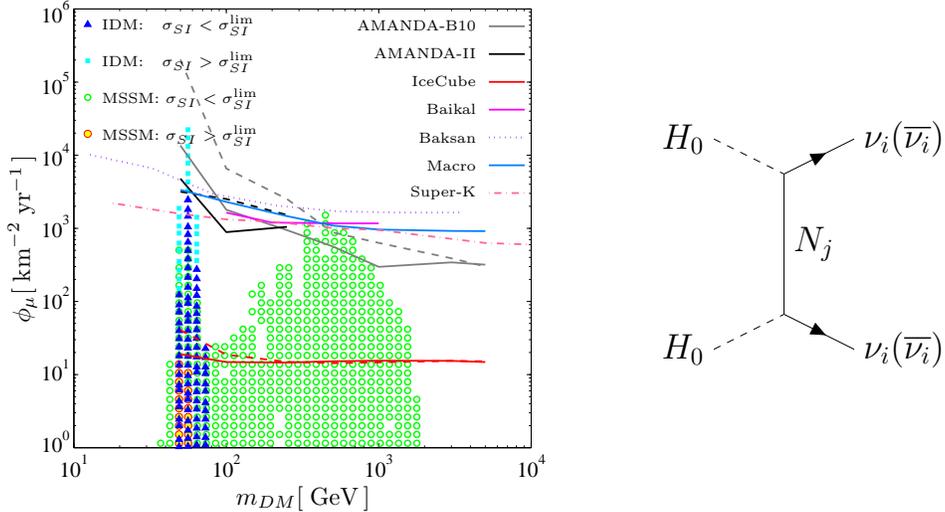}
\caption{\textit{left:} Scatter plot of the muon fluxes from the
Earth for the IDM with Majorana extension (dark blue below, light
blue above direct detection limit) and the MSSM (yellow below, green
above direct detection limit). All plotted points are within the
$3\sigma$ WMAP region. Lines correspond to detector thresholds:
AMANDA-B10 (1997-1999) hard (solid) and soft (dashed); AMANDA-II
(2001-2003) hard (solid) and soft (dashed); IceCube best case hard
(solid) and soft (dashed); BAIKAL (1998-2001); BAKSAN (1978-1995);
MACRO (1989-1998) hard; SUPER-K (1996-2001) soft, see
ref.~\cite{Hubert:2007zza} and references therein. (Parameters: $m_h
= 200 \ \GeV$, $\lambda_2 = 0.2$, $\Delta m_{A_0 H_0} = 8 \ \GeV$,
$\Delta m_{H^+ H_0} = 50 \ \GeV$, $f = 0.3$, $m_N = 100 \ \GeV$, $h
= 0.1$). \newline \textit{right:} Direct annihilation of $H_0$
through exchange of a Majorana neutrino $N_j$ into neutrinos $\nu_i$
(or antineutrinos $\overline{\nu}_i$).} \label{fig-EarthMaj2}
\end{figure}

In figure~\ref{fig-EarthMaj1}, we give the muon flux from
mono-energetic neutrinos for different Higgs masses together with
the IceCube sensitivity for a hard neutrino spectrum (blue lines).
Direct detection constraints (white lines) are the same as in the
previous section, however the WMAP region slightly changes shape due
to the additional annihilation channel into neutrinos. Unless the
Higgs is light (120 GeV) there exists a region of the parameter
space for which the muon fluxes are above the IceCube sensitivity
and agreement with WMAP and direct detection is obtained.

This result is also illustrated in the scatter plot in
figure~\ref{fig-EarthMaj2} which gives the muon flux from the Earth
in function of the dark matter mass together with the sensitivities
of several experiments. The expected muon fluxes for the MSSM are
additionally shown for comparison. The fluxes for the IDM are
typically larger than for the MSSM and lie around the iron resonance
well above the IceCube sensitivity while also agreeing with WMAP and
direct detection constraints. A light Majorana particle as needed
for high fluxes on the negative side gives large one-loop
contributions to the SM neutrino masses ($\sim \keV$) whose
cancelation requires some fine tuning as discussed
in~\cite{Andreas:2009hj}.

\subsection{Sun and light dark matter}

As presented above, the light IDM dark matter candidate has been
studied in~\cite{Andreas:2008xy} in view of explaining the DAMA
annual modulation signature. In this reference, it has been shown
that a large gamma flux from the Galactic centre might be studied
with the Fermi satellite. The most promising neutrino signal however
is expected to come from the Sun which captures light WIMPs well.

\begin{figure}[hbt!]
\includegraphics[clip = true, viewport = 0.7cm 6cm 28.5cm 16.9cm, width=16.4cm]{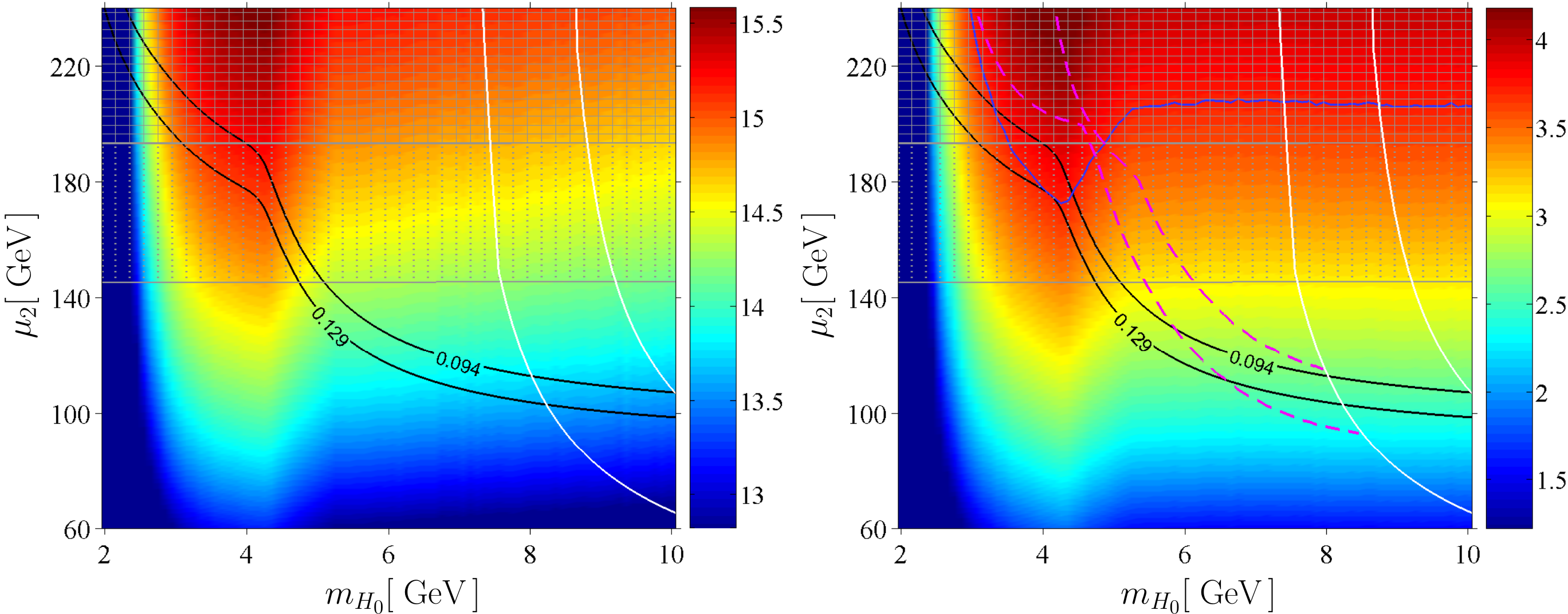}
\caption{Neutrino (\textit{left}) and muon (\textit{right}) fluxes
from the Sun together with our conservative low energy extrapolation
of the Super-Kamiokande sensitivity and the DAMA region in the right
plot. Colour gradient - $\, \log_{10} \phi_\nu ~ [\persqkm \
\peryr]$ \textit{(left)} and $\log_{10} \phi_\mu ~ [\persqkm \
\peryr]$ \textit{(right)}; WMAP area - black lines; DAMA allowed
region - dashed magenta lines; XENON, CDMS limits - white lines,
from left to right; shaded regions correspond to $\lambda_2 > 1$.
(Parameters: $m_h = 120 \ \GeV$, $\lambda_2 = 1$ (dotted) and
$\lambda_2 = \pi$ (shaded),  $f = 0.3$). \label{fig-Sun}}
\end{figure}

In figure~\ref{fig-Sun}, we show both the flux of muon-neutrinos
computed with equation~(\ref{eq-NuFluxSE-CFEQ}) and that of upgoing
muons from equation~(\ref{eq-MuFluxSE}) as a function of the dark
matter mass $m_{H_0}$ and the bare mass scale $\mu_2$. The WMAP
allowed region is given in black and current direct detection
constraints in white (from left to right XENON and
CDMS~\cite{collaboration-2008}). In the right figure of the muon
flux, the DAMA allowed region according to
figure~\ref{fig-DAMA-Gamma} for $f=0.3$ is drawn in addition (dashed
magenta lines).

Since the neutrinos resulting from such a light dark matter
candidate have a quite low energy, they might only be constrained by
the Super-Kamiokande experiment if its sensitivity which currently
reaches down to $m_{DM} = 18 \ \GeV$ is extended to even lower
masses. This might be achieved including stopping, partially and
fully contained muons~\cite{Feng:2008qn}. In figure~\ref{fig-Sun},
we draw in blue an estimate of this sensitivity obtained by
horizontally extrapolating the one given in~\cite{Desai:2004pq} to
lower masses (rescaled to a muon threshold of 1 GeV,
cf.~\cite{Hubert:2007zza}).

It can be seen that the light IDM candidate and a part of the DAMA
allowed region might be tested by Super-Kamiokande but only for a
parameter region in which the coupling $\lambda_2$ is required to be
rather large\footnote{As discussed in~\cite{Andreas:2008xy}, in
order to bring $m_{H_0}$ and $\mu_2$ in the range of WMAP and DAMA
while ensuring the stability of the potential one needs $\lambda_2 >
2 (m_{H_0}^2 - \mu_2^2)^2/(v^2 m_h^2)$.} ($\lambda_2 > 1$ in the
shaded regions). A similar conclusion is presented
in~\cite{Andreas:2008xy}, cf. figure~5 therein, by adapting the
analysis of~\cite{Hooper:2008cf}.

\section{Conclusions}
We presented the signature of the Inert Doublet Model in indirect
detection of neutrinos in three distinct mass ranges and each from a
different place of origin: the Galactic centre, the Earth and the
Sun. For dark matter masses around $1\ \TeV$ we predict an upper
limit on the neutrino flux so that the corresponding gamma flux
stays compatible with the one observed by EGRET. With a boost of
$\mathcal{O}(10^2-10^3)$ from the astro- (dark matter profile)
and/or particle physics (Sommerfeld enhancement) side, this flux
might be observed in Antares. Prospects for the Earth and WIMPs with
a mass $\sim 60 \ \GeV$ are strongly constrained by limits from
direct detection experiments. Even though those WIMPs experience an
enhancement of their capture rate in the Earth due to the iron
resonance, we found no solutions with an observable flux in IceCube
and agreement both with the WMAP abundance and direct detection
limits. This result agrees with the study performed
in~\cite{Agrawal:2008xz}. Assuming the existence of a dark disc in
our Galaxy as indicated by numerical simulations might however give
a sizeable boost $\mathcal{O} (10-100)$ to the predicted fluxes
leading to a more promising prediction. An extension of the model
with light Majorana neutrinos has also been shown to give a
potential signal in IceCube due to the production of mono-energetic
neutrinos via a new annihilation channel. Finally, for the light
dark matter candidate with a mass of a few GeV, we consider the
parameter region that was previously found to be consistent with the
DAMA signal. In agreement with~\cite{Feng:2008dz,Hooper:2008cf}, we
show that neutrinos from the Sun can be used to further test this
scenario with Super-Kamiokande.

\begin{theacknowledgments}
This work has been done in collaboration with Michel H. G. Tytgat.
\end{theacknowledgments}



\bibliographystyle{utphys}   

\bibliography{IDMref}

\end{document}